\documentclass[twocolumn,showpacs,preprintnumbers,amsmath,amssymb]{revtex4}
\usepackage{graphicx}
\usepackage{amsmath}

\begin{document}

\title{ \bf Implementing controlled-NOT gate based on free spin qubits with semiconductor quantum-dot array}
\author{\bf  Yin-Zhong Wu$^{a}$, and Wei-Min Zhang$^{b}$}

\affiliation{ $^{a}$Department of Physics, Changshu Institute of
Technology, Changshu 215500, P.~R.~China\footnote{Email:
yzwu@cslg.cn}\\$^{b}$Center for Quantum Information Science,
National Cheng Kung University, Tainan 70101, Taiwan}

\begin{abstract}

Based on electron spins in semiconductor quantum dots as qubits, a
new quantum controlled-NOT(CNOT) gate is constructed in solid
nanostructure without resorting to spin-spin interactions. Single
electron tunneling technology and coherent quantum-dot cellular
automata architecture are used to generate an ancillary charge
entangled state. Using the ancillary charge entangled state as an
intermediate state, we obtain a spin entangled state and design a
CNOT gate by using only single spin rotations.
\end{abstract}


\maketitle
\section{Introduction}
The idea of using electron spins in semiconductor quantum dots
 as qubits~\cite{book1,PRA1998} has received great attention in the manipulation of a scalabe quantum computer.
Recent experiments~\cite{book2,GaAs}, which show long spin
decoherence time in semiconductor, provide a strong support for
such a picture. It is well known that universal quantum
computation based on electron spins can be achieved by spin-spin
interactions of Heisenberg type\cite{PRA1998,nature1}, XY
type\cite{PRL,PRL1}, and Ising type\cite{Ising}. One may ask
whether it is possible to implement a quantum computation scheme
based on spin qubit without using spin-spin interactions. If it is
possible, then such an idea can supply an alternative approach to
realize quantum computation based on spin qubit. It is the main
motivation of this letter. Very resently, the viewpoint of free
qubit for flying fermions has been proposed~\cite{free}, they
construct a CNOT gate using beam splitter and single spin
rotations if charge detectors are added. However, quantum
computation based on free spin qubit is still lacking in solid
state systems. In this letter, we shall propose an implementation
of a CNOT gate based on solid state nanostructure without
resorting to spin-spin interactions. We use external electrodes to
control single electron tunnelings.  an ancillar charge entangled
state of two electrons is generated with the help of a
multi-quantum-dot structure, i.e., the coherent quantum-dot
cellular automata(CQCA)~\cite{QCQCA}. The charge entangled state
is then converted into a spin entangled state of electrons using
only single spin rotations. Spin-spin interactions are not
required in our scheme, and two-qubit CNOT gate can be easily
manipulated. Thus, a free spin quantum computation can be reached
in semiconductor nanostructure.

\section{Basic Device}
Basic device in our architecture based on a semiconductor
quantum-dot array is shown in Fig.~1. Three quantum dots
(e.g.~1-A-B, 2-C-D) are constructed as a unit cell, and gate
electrodes are integrated in each quantum dot. Solid lines in each
cell indicate the possibility of interdot tunneling. The
confining barrier between two cells must satisfy that the
tunneling of electrons between different cells is forbidden.
Moreover, there exists only one excess conductor electron in each
cell. Spin states of the excess electron are chosen as qubit
states.
\begin{figure}[ht]
   \centering
   \includegraphics[width=3.0in]{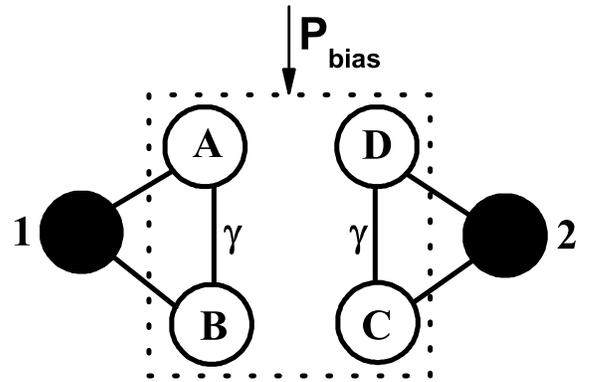}
   \caption{Architecture of our CNOT gate composed of two unit
cells(each cell contains three QDs). Four quantum dots of
neighboring unit cells (e.g. ABCD) form a coherent quantum-dot
cellular automata structure(the dotted square box).
 $\gamma$ stands for the tunneling energy of electron between
vertical QDs, and $P_{bias}$ denotes the bias polarization.}
\end{figure}
Meanwhile, four quantum dots of neighboring unit cells (e.g.~ABCD)
form a usual coherent quantum-dot cellular automata(CQCA)(the
dotted square box in Fig.~1). In order to give a clear picture of
CQCA, let's describe quantum-dot cellular automata(QCA) at first.
Metal QCA has been realized in experiment to simulate the
classical digital algorithm\cite{science2}. A semiconductor
realization of QCA structure has also been developed\cite{SQCA}.
When the QCA structure is charged with two excess electrons, two
electrons will occupy diagonal sites as a result of Coulomb
repulsion. Moreover, only the full polarization (diagonal
polarization) charge states can be existed in QCA structure. The
structure of CQCA is first proposed by G.~Toth\cite{QCQCA}.
Different from the charge polarization states in QCA, the charge
state of two electrons in CQCA can be an arbitrary superposition
state of the two full polarizations if coherent operations are
performed on the full polarization states.  Two input
parameters($\gamma$ and $P_{bias}$) are introduced to CQCA
structure~\cite{QCQCA}, where $\gamma$ is the tunneling energy
between vertical QDs in CQCA, $P_{bias}$ is the bias polarization
which can be implemented by bias gate voltages. The effective
interaction between the external bias polarization and the charge
state of two electrons within CQCA can be expressed as
$E_{0}P_{bias}\sigma_{z}$\cite{QCQCA}, where $E_{0}$ is the
strength of Coulombic coupling of two electrons, and the eigen
states of $\sigma_{z}$ are the two full polarization states. We
will show later that only a relative phase between the two full
polarization states is required in our scheme. So, it is not
necessary to tune the tunneling energy between vertical QDs in
our CNOT gate. Turning on $P_{bias}$, the relative phase factor
will be reached if $E_{0}P_{bias}\gg \gamma$. Thus, only one input
parameter $P_{bias}$ in CQCA is needed to implement our CNOT gate.

\section{Definition of qubit states}

The state of the unique electron in unit cell $i$ is defined as a
direct product of electron charge and spin state $|e_{i}\rangle |
S_{i}\rangle $($i=1,2$). We take electron spin states as qubit
states. At the initial time, the excess conductor electrons site
in quantum dot 1 and 2. The excess electron in each cell is driven
away from its initial position only when a two-qubit operation is
performed, and electrons will be pushed back to their initial
positions as soon as a two-qubit operation is completed. So, we
define the electron position in unit cell $i$ at the initial time
as the charge "ground state" $|e_{g_{i}}\rangle$.
$|e_{1}^{A}\rangle$ and $|e_{1}^{B}\rangle$ are defined as the
charge state of electron in the left unit siting at QDs A and B,
respectively, while $|e_{2}^{C}\rangle$ and $|e_{2}^{D}\rangle$
are defined as the charge state of electron in the right unit
siting at QDs C and D, respectively. Control of the location of
the excessive electron within each unit cell can be realized by
turning on/off the gate voltage of each QD.

\section{Implementation of a CNOT operation}
Four steps are needed to implement a CNOT operation on the two
qubits within two neighboring unit cells.
 Firstly, turning on the gate voltage to lower the site-energy in
 dots A, B, C, and D, the electron in QD $1$ will tunnel into QDs $A$ and $B$. If QDs $A$ and $B$
 are identical, the probabilities of electron $1$
tunneling into QDs $A$ and $B$ are equal. In some sense, this
process has some analogy to a flying electron moving through a
beam split except that here the case is based on solid system. At
the same time, electron $2$ tunnels into QDs $C$ and $D$.
According to the special property of cellular automata
structure(Coulomb repulsion principle), two electrons will
automatically occupy on the two diagonal positions in the four
QDs(AC and BD) with equal probability. The state of the two
electrons before and after this process can be written as~\cite{cq}
\begin{equation}
|e_{g_{1}}\rangle|e_{g_{2}}\rangle
|S_{1}\rangle|S_{2}\rangle\Longrightarrow
\frac{1}{\sqrt{2}}(|e_{1}^{A}e_{2}^{C}\rangle+|e_{1}^{B}e_{2}^{D}\rangle)|S_{1}\rangle|S_{2}\rangle.
\end{equation}
Then, applying bias gate voltages on electrodes in CQCA, when the
duration of the bias polarization satisfies
$\int_{0}^{T}{E_{0}P_{bias}(t) dt}=\frac{\pi}{4} $, a specific
relative phase is realized, and an ancillary charge entangled
state, i.e.,
$\frac{1}{\sqrt{2}}(-i|e_{1}^{A}e_{2}^{C}\rangle+|e_{1}^{B}e_{2}^{D}\rangle)$,
can be obtained, where a global phase $e^{i\frac{\pi}{4}}$ has
been discarded. As mentioned before, the strength of
$E_{0}P_{bias}$ should be much larger than the tunneling energy
$\gamma$. Quantum computation using two full polarization charge
states in a coherent quantum-dot cellular automata as qubit states
has been proposed\cite{QCQCA}.
 Different from their works, here we
choose electron spin states as qubit states, and the quantum-dot
cellular automata is used only as an ancillary structure. So far,
we did nothing upon electron spin states(qubit states). In the
following steps, based on the ancillary charge entangled state, we
will present how to implement a CNOT operation with only single
spin rotations.

Secondly, two single spin rotations $R_{Z}^{A}(3\pi)$ and
$R_{X}^{C}(\pi)$ are performed on QDs $A$ and $C$, respectively,
where $R_{a}(\theta)$ is defined as $e^{-i\theta a/2}$. Note that
the two single spin operations can be executed simultaneously. The
single spin rotations can be implemented by either a local
magnetic field or ultrafast optical pulses as we shall discuss
later. After the second step, the state of the two excess
electrons becomes
\begin{equation}
\frac{1}{\sqrt{2}}(-i|e_{1}^{A}e_{2}^{C}\rangle
R_{Z}^{A}(3\pi)|S_{1}\rangle
R_{X}^{C}(\pi)|S_{2}\rangle+|e_{1}^{B}e_{2}^{D}\rangle|S_{1}\rangle|S_{2}\rangle).
\end{equation}
The third step is an inverse process of the first step. Tuning the
gate voltage inversely, electron $1$ and electron $2$ will return
their original locations, and the ancillary charge entangled state
collapses into its ground state
$|e_{1}^{A}e_{2}^{C}\rangle\longrightarrow
|e_{g_{1}}\rangle|e_{g_{2}}\rangle$,
$|e_{1}^{B}e_{2}^{D}\rangle\longrightarrow
|e_{g_{1}}\rangle|e_{g_{2}}\rangle$. Spin entangled state is
generated through the collapse of the ancillary charge entangled
state, and the state of the two electrons changes to
\begin{eqnarray}
&&|e_{g_{1}}\rangle|e_{g_{2}}\rangle\frac{1}{\sqrt{2}}(-iR_{Z}^{A}(3\pi)|S_{1}\rangle R_{X}^{C}(\pi)|S_{2}\rangle+|S_{1}\rangle|S_{2}\rangle)\nonumber\\
&&=|e_{g_{1}}\rangle |e_{g_{2}}\rangle
e^{-i\frac{\pi}{4}\sigma_{z}\otimes\sigma_{x}}|S_{1}\rangle|S_{2}\rangle,
\end{eqnarray}
 explicitly, after the third step, we write down the state of the
two qubits for different initial states $|00>,|01>,|10>$, and
$|11>$ as following:
\begin{eqnarray}
 &&|00>  \rightarrow  \frac{1}{\sqrt{2}}(|00>-i|01>),\nonumber\\
 &&|01> \rightarrow  \frac{1}{\sqrt{2}}(-i|00>+|01>), \nonumber\\
&&|10>   \rightarrow  \frac{1}{\sqrt{2}}(|10>+i|11>), \\
&&|11>  \rightarrow \frac{1}{\sqrt{2}}(i|10>+|11>).\nonumber
\end{eqnarray}
Fourthly, we do another two single qubit rotations
$R_{Z}^{1}(3\pi/2)$ and $R_{X}^{2}(3\pi/2)$ on QDs $1$ and $2$,
respectively. We note that the two single spin operations can also
be performed at the initial time. When the four steps are
completed, we get a CNOT operation acting on two qubits within the
basic device, namely,
\begin{eqnarray}
 &&|00> \rightarrow |00>,\nonumber\\
 &&|01> \rightarrow |01>,\nonumber\\
 &&|10> \rightarrow  |11>,\\
 &&|11> \rightarrow |10>.\nonumber
\end{eqnarray}

In our scheme, single qubit rotation can be realized via a local
magnetic field or ultrafast laser pulses. Single spin rotations
$R_{X}$ and $R_{Z}$ can be implemented by turning on local
magnetic fields along $x$ direction and $z$ direction,
respectively. However, one-qubit operation by applying a magnetic
field is very slow. For speeding up the operations, we shall use
two laser pulses to generate a single spin rotation through
adiabatic process\cite{adiabatic} or use optical "tipping"
pulse\cite{tip}. In the following, we will describe how to realize
a single spin rotation by using Raman transition with adiabatic
process. Two pulsed lasers with different polarizations are
addressed on a quantum dot. For a large detuning and under Raman
resonance($\Delta=\Delta_{1}$=$\Delta_{2}\gg \Omega_{eff}$), i.e.,
adiabatically eliminating the excited state\cite{adiabatic}, an
effective interaction between spin-up state $|0\rangle$ and spin
down state $|1\rangle$ can be realized:
$\Omega_{eff}(t)(|0\rangle\langle
1|e^{-i[(\omega_{1}-\omega_{2})t+\delta\phi_{12}]}+|1\rangle\langle
0|e^{i[(\omega_{1}-\omega_{2})t+\delta\phi_{12}]})$, where
$\Omega_{eff}(t)=\frac{\Omega_{1}(t)\Omega_{2}(t)}{\Delta}$,
$\delta\phi_{12}$ is the initial phase difference between the two
lasers, $\Delta_{1}$ and $\Delta_{2}$ denote the detuning of two
lasers, respectively, and $\Omega_{1}$($\Omega_{2}$) denotes Rabi
frequency between spin-up state $|0\rangle$(spin-down state
$|1\rangle$) and the excited state of the quantum dot. By properly
adjusting the duration and the initial phase of each laser pulse
to satisfy $\int_{0}^{T}\Omega_{eff}(t)dt=\frac{\theta}{2}$ and
$(\omega_{1}-\omega_{2})T+\delta\phi_{12}=2n\pi$, a single spin
rotation  $R_{X}(\theta)$ is achieved, and $R_{Y}(\theta)$ could
be implemented under the conditions
$\int_{0}^{T}\Omega_{eff}(t)dt=\frac{\theta}{2}$ and
$(\omega_{1}-\omega_{2})T+\delta\phi_{12}=2n\pi+\frac{3\pi}{2}$.
Combining $R_{X}(\theta)$ with $R_{Y}(\theta)$, we can obtain an
arbitrary single spin rotation $R_{Z}(\theta)$. For example,
single spin rotation $R_{Z}(3\pi)$ performed on QD $A$ can be
achieved by $R_{X}(3\pi)R_{Y}(\pi)$, and single spin rotation
$R_{Z}(\frac{3\pi}{2})$ performed on QD $1$ can be realized by
three series operations
$R_{Y}(\frac{3\pi}{2})R_{X}(\frac{\pi}{2})R_{Y}(\frac{\pi}{2})$ .
In principle, QDs $1$, $2$, $A$ and $C$ can be different in size.
Masks might be manufactured upon QDs $B$ and $D$\cite{mask}, then
the laser pulse can selectively couple to an individual QD, and
the laser field need not be localized spatially in our scheme.

As the initial state preparing is concerned, bias gate voltages
can be used to lower the on-site energy of dots $1$ and $2$ for
the initialization of charge state, and a static uniform magnetic
field can be applied to split spin-up state and spin-down state by
the Zeeman energy for the initialization of spin state(qubit
state).

Recently, single electron tunneling through triple virtual
quantum dots has been studied in experiment\cite{APL}. Although
our scheme would be a challenge in technology, it gives a new way
on how to implement free spin quantum computation based on solid
QDs array. Such structures could be fabricated in scalable, and we
envisage that the separation between each CNOT gate structure
would be greater than the size of them. This ensure that the
Coulomb repulsion between electrons in different CNOT gates is
always smaller than the Coulomb repulsion between two electrons
within the same cellular automata structure and may be treated as
perturbation. As decoherence time and operation time are
concerned, the typical decoherence time of electron spin in GaAs
QD is $50\mu s$\cite{GaAs}, the operation time of single spin
rotation by ultrafast pulse is several picoseconds, and the time
of single electron tunneling between two coupled QDs is about a
few 100 ps\cite{tunnel}. Thus, the great challenge in our scheme
might be the fast control technology of SET\cite{SET}. Here, We
focus our attention on how to design a CNOT gate, the architecture
for scalable quantum computation, measurement, and how to use this
basic device to produce spin Bell states will be discussed
elsewhere\cite{zws}. We hope the device can be fabricated by
current or future nano-technology, and the CNOT gate might be
tested firstly in experiment before a scalable quantum computation
is put forward.

\section{Conclusion}

In summary, an ancillary charge entangled state is generated based
on semiconductor coherent quantum-dot cellular automata. Since
spin and charge are commute, the spin entangled state is realized
based on the ancillary charge entangled state. A CNOT gate based
spin qubits is first implemented in solid system without using of
qubit-qubit interactions. Combining with single qubit operations,
a scalable free spin quantum computation could be realized in
semiconductor nanostructure.



\begin{thebibliography}{99}
\bibitem{cq}As described by Eq.~(6) in Ref.~\cite{QCQCA}, when two parameters
$\gamma$ and $P_{bias}$ are introduced into the QCA structure, one
can obtain an arbitrary superposition state of the two diagonal
polarization states. One can also abtain Eq.~(1) by employing the
following two steps. Firstly, two electrons in two unit cells are
driven to one of the diagonal polarization state. Secondly, by
applying coherent operations ($\sigma_{X}$, $\sigma_{Z}$) on the
diagonal polarization state, the charge entangled state, described
by Eq.~(1), can be realized.
\end{thebibliography}
\end{document}